\documentclass{aa}  

\usepackage{graphicx}
\usepackage{subcaption}
\usepackage{physics}
\usepackage{xcolor}
\usepackage{float}
\usepackage[normalem]{ulem}
\usepackage[switch]{lineno}
\usepackage{amsmath}
\usepackage{placeins}

\usepackage{txfonts}
\usepackage[hidelinks]{hyperref}
\usepackage{cleveref}
\crefname{figure}{Fig.}{Figs.}

\newcommand{\ha}{{H$\alpha$ }}

\begin{document} 

\title{Fast Bayesian spectral analysis using Convolutional Neural Networks: Applications over GONG \ha solar data}

\subtitle{}

\author{G. Castelló
      \inst{1,2}
      \and
      M. Luna 
      \inst{1,2}
      \and
      J. Terradas
      \inst{1,2}
      }

\institute{Departament de Física, Universitat de les Illes Balears (UIB), E-07122 Palma de Mallorca, Spain\\
          \email{guillem.castello@uib.cat}
     \and
         Institute of Applied Computing \& Community Code (IAC3), UIB, E-07122 Palma de Mallorca, Spain}

\date{Received November 08, 2024; accepted January 17, 2025}


\abstract
{
Solar filament oscillations have been observed for many years, but recent advances in telescope capabilities now enable daily monitoring of these periodic motions, offering valuable insights into the structure of filaments.
A systematic study of filament oscillations over the solar cycle can shed light on the evolution of the prominences. Until now, only manual techniques have been used to analyze these oscillations.}
{
%
This work serves as a proof of concept, aiming to demonstrate the effectiveness of Convolutional Neural Networks (CNNs). These networks automatically detect filament oscillations by applying 
power-spectrum analysis to \ha data from the GONG telescope network.
}
{The proposed technique studies periodic fluctuations of every pixel of the \ha data cubes. Using the Lomb Scargle periodogram, we computed the power spectral density (PSD) of the dataset. The background noise is well fitted to a combination of red and white noise. Using Bayesian statistics and Markov chains - Monte Carlo (MCMC) algorithms we can fit the spectra and determine the confidence threshold of a given percentage to search for real oscillations. We built two CNN models to obtain the same results as the MCMC approach.}
{We applied the CNN models to some observations reported in the literature, proving its reliability in detecting the same events as the classical methods. A day with events not previously reported has been studied to check for the model capabilities outside of a controlled dataset where we can check with previous reports.}
{CNNs proved to be a useful tool to study solar filament oscillations, using spectral techniques. Computing times have been reduced significantly while getting results similar enough to the classical methods. This is a relevant step towards the automatic detection of filament oscillations.}

\keywords{GONG H$\alpha$. Solar filament. Spectral Bayesian analysis. Markov chain Monte Carlo (MCMC). Convolutional neural network (CNN).}

\maketitle
%

\section{Introduction}
Solar filaments—also known as prominences when observed off-disk—are dense, cool plasma structures suspended in the Sun's hot corona by magnetic fields, appearing as dark 
structures in \ha observations. Despite extensive research, the formation, structure, and disappearance of these filaments are not fully understood. They form within filament channels (FCs), which are large, non-potential magnetic structures situated above the polarity inversion lines (PILs) in the photosphere. The development of FCs is influenced by solar surface and subsurface motions, including differential rotation, meridional flows, and convection \citep[see, e.g.,][]{mackay2014formation}.
Throughout the solar cycle, filaments migrate across different latitudes, forming patterns that resemble the sunspot butterfly diagram \citep{1972RvGSP..10..837M,1998SoPh..177..357M,1998ASPC..150..488C, diercke_universal_2024}. This migration suggests a relationship between filament structures and the large-scale solar magnetic field. 

Solar prominences exhibit various types of motion, notably oscillations, which have become more frequently observed due to the capabilities of new instruments \citep{arregui_prominence_2018}. These oscillations are relevant because they enable prominence seismology to determine physical parameters of magnetized plasma structures.
Oscillations in solar filaments are categorized based on their restoring forces. Transverse oscillations are mainly governed by the Lorentz force \citep{1966ZA.....63...78H, 1969SoPh....6...72K, 2011A&A...531A..53H}, while longitudinal oscillations are influenced by gravity along magnetic field lines and gas pressure gradients \citep{luna_large-amplitude_2012,luna_effects_2012,2012A&A...542A..52Z, zhang_parametric_2013}. \citet{2017ApJ...850..143L} compared field properties derived from filament seismology with the reconstructed magnetic field using the flux rope insertion method by \citet{2004ApJ...612..519V} and found a general agreement. This highlights the potential of oscillation studies to enhance our understanding of filament structures.

Despite the potential insights that systematic studies of filament oscillations could offer into the evolution of prominences and their FCs over the solar cycle, most observational research has focused on individual events. 
One of the few extensive studies was conducted by \citet{1993A&A...279..610B}, who reported 38 oscillation events over nearly a solar cycle and noted a sinusoidal latitudinal dependence of oscillation periods. However, a later study by \citet{luna_gong_2018} did not find a clear relationship between oscillation periods and filament latitude. Using \ha data from the GONG network during January–June 2014, \citeauthor{luna_gong_2018} provided an extensive sample of 196 events near the solar maximum of cycle 24 and found an average oscillation period of 58 minutes, regardless of amplitude or filament type.

Traditionally, the time-distance technique has been used, which involves measuring the intensity along a fixed artificial slit following the motion. This method has been widely applied in the study of oscillations in coronal loops and the detection of solar filament oscillations \citep[see][]{pant_simultaneous_2016, zhang_simultaneous_2017, joshi_interaction_2023}.
%
However, these methods require firstly visual detection of the oscillations and secondly tracing an artificial slit to construct the time-distance diagram.
This makes time-distance diagram methods not applicable for the automatic detection and parameterization of oscillations in filaments.
\citet{luna_automatic_2022} introduced a new method by using two-dimensional \ha images and performing spectral (frequency-domain) analysis on the intensity time series of all pixels in the data. This technique allows the automatic detection of oscillations in the solar disk.
%

%
\citet{luna_automatic_2022} demonstrated that spectral techniques can effectively detect oscillations in solar filaments, serving as a proof of concept. However, the methods they employed offer room for refinement. Following the approach of \citet{auchere_fourier_2016}, background noise is typically estimated by fitting the PSD using a weighted least squares method. This fit is weighted by the inverse square of the smoothed PSD, obtained via a five-point boxcar average. Additionally, confidence levels are calculated assuming that the background noise is white noise.
The PSD derived from the temporal evolution of pixel intensities is usually a combination of red and white noise. Analyzing such PSDs is challenging because red noise can mimic or obscure genuine oscillatory signals, potentially leading to false detections if not properly accounted for. Sophisticated statistical methods are therefore necessary to distinguish true oscillations from noise-induced fluctuations. While some studies have attempted to simplify the analysis by whitening the PSD to mitigate the red noise component, \citet{vanderplas_understanding_2018} emphasizes that a more rigorous approach involves following the Bayesian methodology outlined by \citet{vaughan_bayesian_2010}. This methodology employs MCMC techniques to assess the significance of detected oscillations robustly.

Although these Bayesian methods offer a statistically robust framework, their computational complexity can be prohibitive for the large datasets typical in solar imaging. In our work, we initially employed the MCMC-based approach but found it too slow for our research purposes. To overcome this limitation, we decided to use machine learning. In recent years the field of machine learning has bloomed in popularity in many scientific fields, like in solar physics  \citep[see][]{asensio_ramos_machine_2023}, for the benefits it provides in computational efficiency, and the ability to process enormous amounts of data. For this reason, we developed a deep learning model replicating the MCMC method's predictive capabilities, substantially improving computational efficiency without compromising accuracy.

The paper is structured as follows. In Section 2 the \ha data from the GONG telescope network is described. Section 3 describes the theoretical background for oscillation identification in red-noise PSD using the methods described in \citet{vaughan_bayesian_2010}. Section 4 presents the deep learning model which uses CNNs as its architecture. Section 5 shows the results for three different observational days, two of which have events previously reported in \citet{luna_gong_2018} and one not previously reported by other authors. Finally, in Section \ref{sec:discussion}, the results are discussed and conclusions are drawn.


%
%
%
\section{Data}
In \ha images, filaments appear as dark structures against the bright chromospheric background, which makes them particularly useful for studying filament dynamics. This clarity is often lacking in EUV data from instruments like AIA/SDO \citep{lemen_atmospheric_2011}, where the more complex and dynamic structures complicate filament identification and the automated detection of their oscillations.

\subsection{NSO-GONG network}
The NSO-GONG\footnote{\url{http://gong2.nso.edu}} \citep{hill_global_1994} telescope network monitors the full disk almost continuously in the \ha wavelength
The network consists of telescopes with identical design and construction, placed around the globe at the following locations: Learmonth (L), Udaipur (U), El Teide(T), Cerro Tololo (C), BigBear (B), and Mauna Loa (M). These locations were strategically selected to follow the diurnal motion of the Sun, to ensure full-day coverage. 
Each telescope in the network captures images during its daytime, with overlapping hours of coverage between consecutive telescopes following the Earth’s rotation.
Since the NSO-GONG network is ground-based, Earth's atmospheric conditions can influence the observations and should be considered. The \ha GONG data consists of images of 2048$\times$2048 pixels with a 
nominal spatial resolution of approximately 
2 arcsec. Each of the network's telescopes provides images with a one-minute cadence. The network's duty cycle is 90\% meaning that due to telescope performance or poor sky quality, eventually, the telescope does not provide the image resulting in data gaps in the time series. 
%

\subsection{Data preparation}\label{sec:data-preparation}

In this work, we analyze the temporal sequences of GONG \ha images. 
%
%
We have grouped the GONG data into full-day bins. 
Although this choice is arbitrary, it enables the grouping of detections by day.
%
The choice of the length of time sequences also imposes a limit on the maximum period that can be detected.
However, oscillations primarily exhibit periods of less than a few hours \citep[see, e.g.][]{arregui_prominence_2018,luna_gong_2018}. In the future, longer bins spanning several days will be used to explore oscillations of longer periods, such as those observed by \citet{foullon_detection_2004,foullon_ultra-long-period_2009} and \citet{efremov_ultra_2016}.
%
%
%

To construct the \ha time series, it is necessary to combine data from the various telescopes in the GONG network. Each GONG telescope operates for several hours throughout the day, with the possibility of up to four telescopes operating simultaneously.
During these overlaps it becomes necessary to select which telescope's data to use for constructing the time series. To facilitate this selection process, we employ the following algorithm:
\begin{enumerate}
\item We first segment the full-day time series into intervals where data is either absent, available from a single telescope, or overlapping from multiple telescopes. In cases where only one telescope is active, we use its data by default.
\item When two or more telescopes are observed within the same time window, we evaluate the quality of the images produced by each telescope. The telescope providing the highest quality images is then selected for that segment. 
\item Once the image sequence is constructed, we remove limb darkening from the images \citep{cox_allens_2002}. Differential rotation is then compensated by remapping all images to 12:00 UT using SunPy\footnote{ \url{https://docs.sunpy.org/en/stable/topic_guide/coordinates/rotatedsunframe.html}} algorithms. This ensures consistent intensity across the entire disk, allowing each pixel to correspond to the same area on the solar surface throughout the temporal sequence.
\end{enumerate}
The selection of the appropriate telescope in step 2 is guided by a preference for longer uninterrupted sequences from a single telescope to minimize the frequency of switching between telescopes. Additionally, we consider the average sharpness value of the image sequence as recorded in the FITS metadata to ensure optimal image quality.

After this process, individual images that may have poor quality or atmospheric noise, such as cloud cover, are checked for exclusion. In some cases, only a portion of the solar disc is visible due to issues with the telescope.
These images generate rapid intensity variations, resulting in significant high-frequency noise.
To identify these problematic images, quick checks of their circular symmetry relative to the image center are performed.
We calculate Pearson's correlation coefficient between the \ha intensities along the central horizontal and vertical lines, denoted as $\rho_\oplus$, as well as along the two main diagonals, denoted as $\rho_\otimes$. In high-quality images, both $\rho_\oplus$ and $\rho_\otimes$ should be similar and close to one, even when dark structures like filaments are present.
To assess image quality for a day of observations we proceed as:
\begin{enumerate}
    \item Compute $\rho_\oplus$ and $\rho_\otimes$ for each image within the corrected temporal sequence.
    \item Calculate the mean values, $\overline{\rho}_{\oplus}$ and $\overline{\rho}_{\otimes}$, and standard deviations, $\sigma_{\rho_\oplus}$ and $\sigma_{\rho_\otimes}$, of these coefficients over the entire temporal sequence.
    \item All images that have either $\rho_\oplus$ or $\rho_\otimes$ below the threshold defined as $\overline{\rho_{\oplus , \otimes}} - 2 \, \sigma_{\rho_{\oplus , \otimes}}$ is considered to have poor quality and discarded
\end{enumerate}

\begin{figure}[!ht]
   \resizebox{\hsize}{!}
            {\includegraphics[width=0.5\textwidth]{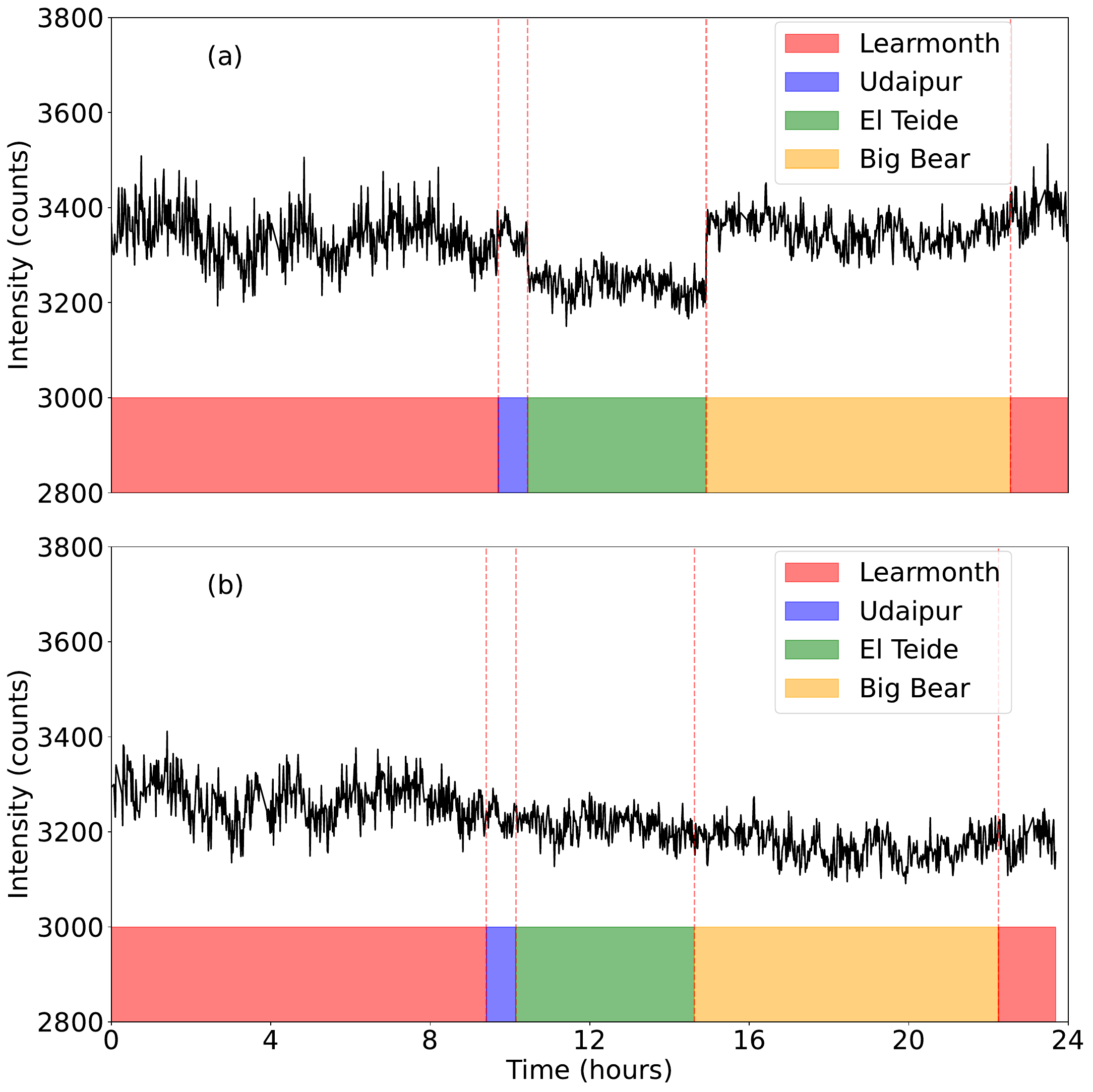}}
    \caption{Observation windows from the GONG telescope network for a random pixel during 22 October 2024. Red vertical dashed lines represent the times when there has been a telescope change. (a): Raw time series without any preprocessing applied over it. (b): Preprocessed time series.}
    \label{fig:obs_windows}
\end{figure}

Figure \ref{fig:obs_windows} shows the effects of the different pre-processing steps we applied. We plotted the temporal series of a given pixel on the 22 October 2024. In \cref{fig:obs_windows}a the full temporal sequence is shown with 1418 candidate images.
The limb-darkening and differential rotation have been corrected in these images.
%
The image quality evaluations have been applied and 29 images have been removed from the original set and the result is plotted in \cref{fig:obs_windows}b.
%
%
We have checked that this technique discards low-quality images satisfactorily and quickly.
Nowadays, there are methods to remove clouds from Earth's atmosphere, based on deep learning techniques as by  \citet{wu_algorithm_2021} and \citet{chaoui_removing_2023}. However, we prefer to remove images containing clouds as they are a relatively small fraction of the full-day data. In addition, it should be ensured that these techniques do not introduce artifacts in the intensity temporal series. We will consider these techniques in future studies.

With the methods described above, we obtain a sequence of data by combining the different telescopes and removing poor-quality images.
%
Due to local atmospheric conditions, the Sun’s position above the horizon, slight differences in sensitivities, etc., the intensities measured by the different network telescopes vary slightly. Figure \ref{fig:obs_windows}a shows the full-day temporal sequence of the intensity in a given pixel of the day 22-10-2024 of 1418 images. There is an abrupt drop in intensity in the El Teide data with respect to Udaipur and Big Bear. 
To ensure continuity and avoid intensity jumps during telescope changes, we 
apply a pixel-wise correction of the intensity differences by computing the median of the last 10 intensity values from telescope $n$ and the first 10 intensity values from telescope $n+1$ each time a swap occurs at time $t_n$.
Then, we compute the difference between these two medians and add it to the entire temporal series of telescope $n+1$. The values of telescope $n+1$ are updated, and this operation is repeated for each telescope swap. In this way, a continuous intensity sequence is obtained, as shown in \Cref{fig:obs_windows}b.
The next step is the correction of any outlier pixel intensities that can be present in the corrected temporal data. It is done as follows:
\begin{enumerate}
    \item Take a pixel temporal series and compute the moving median with a kernel size of 7.
    \item 
    Subtract this moving median to the original signal, take the absolute value and compute the standard deviation of the result.
    \item If at any time instant the "normalized" time series is above four times the obtained standard deviation, set that value in the original data to the median at that instant.
    \item Repeat for all the other pixels.
\end{enumerate}
Finally, the last step is to set all the pixels outside the solar disk equal to zero, to remove any signals from regions outside of the disk.


\section{Spectral analysis: identifying oscillations}\label{sec:spectral-analysis}

After processing the GONG data we obtain sequences of images grouped by day through the combination of different telescopes. The time series obtained are normalized and apodized to remove high-frequency noise. This pre-processing results in a data cube that is subsequently analyzed as explained in this section.
We analyze the temporal sequences in each of the pixels of the images by constructing PSD. For each day, there is a maximum of 1440 data points in the temporal sequence. However, there can be gaps in the data, resulting in irregular sampling.
For this reason, we use the generalized Lomb-Scargle periodogram \citep[LS-periodogram;][]{lomb_least-squares_1976, scargle_studies_1982, carbonell_search_1991, zechmeister_generalised_2009} to generate the PSD. The term ``generalized’’ indicates that the signal does not need to be zero-averaged, unlike in the original versions of the periodogram \citep[see][for further details]{vanderplas_understanding_2018}.
%
The resulting PSD is normalized following the standard normalization approach outlined in the Astropy documentation \footnote{  \url{https://docs.astropy.org/en/stable/timeseries/lombscargle.html}} which results in a dimensionless PSD, with values ranging from 0 to 1.

Identifying periodic signals consists of discerning real oscillations among noise fluctuations. It is necessary to model the background noise power and determine the confidence level. The background noise power is the PSD of a signal in the absence of real oscillations.
The background noise in the GONG data can be modeled as a combination of red and white noise \citep{luna_automatic_2022}.
For each pixel of the \ha images time series we consider the background noise power as 
\begin{equation}\label{eq:backgroundnoise} 
    S(\nu) = A\, \nu^{-\alpha} + B \, ,
\end{equation}
where $\nu$ is the frequency and $A, \alpha$ and $B$ are free parameters. The first power-law term models the red noise whereas the second constant term is the white noise. 
The red-noise nature of the PSD adds a layer of complexity to the analysis.
\citet{vaughan_simple_2005} proposed a method for analyzing the red noise section of the spectrum separately from the white noise part. Using such technique \citet{gruber_quasi-periodic_2011} showed that applying a white-noise assumption to data with a power law PSD drastically overestimates the significance of the detected peaks.
Another common way to deal with such data relied on subtracting the "background components" of the time series (i.e. moving average). \citet{vaughan_bayesian_2010} showed that this procedure may lead to misleading results and should be avoided; the entire Fourier spectrum of the signal should always be considered.
For these reasons, we use the methods of Bayesian statistics, employing the same techniques as described in \citet{vaughan_bayesian_2010} and \citet{inglis_quasi-periodic_2015}.
%

\citet{vaughan_bayesian_2010} showed the theoretical foundation for parameter estimation in red-noise data using Bayesian statistics and MCMC methods.
%
In general, in this method, given the data $\textbf{D} = \{D_1, \ldots, D_N\}$, a model $H$, and a set of parameters $\theta = \{\theta_1, \ldots, \theta_M\}$, we aim to determine the probability of these parameters given the data and our model, expressed as $p(\theta|\textbf{D}, H)$ which is referred as the posterior distribution. We make use of Bayes' Theorem, namely
\begin{equation}\label{eq:bayes-theorem}
    p(\theta|\textbf{D}, H) = \frac{p(\textbf{D}|\theta, H) p(\theta | H)}{p(\textbf{D}|H)} \, ,
\end{equation}
where 
$p(\textbf{D}|\theta, H)$ is the likelihood of the parameters  (the probability of the data occurring given that our model and parameters are true), $p(\theta | H)$ is the prior probability (the information we have beforehand of the parameters) and finally $p(\textbf{D}|H)$ is known as the marginal likelihood (of the data) or the prior predictive distribution that is just a normalization constant. 
%

In this study, the data $\textbf{D}$ consist of the PSD, where $D_j = \mathrm{PSD}(\nu_j)$ in Eq. \eqref{eq:bayes-theorem}. The model $H$ is given by Eq. \eqref{eq:backgroundnoise} and parameters $\theta$ are $A$, $\alpha$, and $B$.
The null hypothesis is that our data comes from a stochastic process, so the periodogram 
is distributed around the background noise ($S_j = S(\nu_j))$, following a $\chi_2^2$ distribution. With this hypothesis, we can write our likelihood function as:
\begin{equation}
    p(\textbf{D}|\theta, H) = \prod_{j} \frac{1}{S_j} exp\bigg( -\frac{D_j}{S_j} \bigg) \, .
\end{equation}
Our prior distributions are uniform probability functions, chosen to reflect our initial lack of specific knowledge about the parameter values, with all values within the specified ranges considered equally probable. We intentionally selected broad ranges for these priors to avoid imposing restrictive assumptions on the parameter space. The chosen ranges are
\begin{equation}
   \begin{array}{l}
      0 < A < 1 \, ,\\
      0 < \alpha < 10 \, ,\\
      0 < B < 1 \, .
   \end{array}
\end{equation}
These broad prior ranges are selected to encompass the range of likely parameter values while preserving flexibility in our model.

The marginal likelihood $p(\textbf{D}|H)$ is a normalization constant that is computed by integrating over the whole parameter space the product of the prior probability function and likelihood function. This integral is often difficult to compute exactly and may not have an analytical solution. 
To avoid its direct computation, MCMC algorithms are used to obtain the posterior distribution. 
After an initial phase called the ``burn-in'' period, the Markov chain generates samples of the parameters $\theta$ that are distributed according to the Bayesian posterior probability density. In other words, once equilibrium is reached, the distribution of the Markov chain’s samples mirrors the posterior probability density function (PDF). 
Consequently, regions of the parameter space corresponding to higher posterior probabilities are sampled more frequently than those with lower probabilities. Therefore, by taking enough samples, we can ensure that we accurately model the posterior PDF for the parameters.
The posterior PDF contains valuable information: the mode of the parameters $\theta_{mode}$ provides the best fit to the data with the proposed model. Additionally, with the posterior distributions the $T_R$ test statistic can be computed:
\begin{equation}\label{eq:tr}
    T_R = \max_j (R_j) = \max_j (2 D_j / S_j ) \, ,
\end{equation}
where $R_j = \frac{2 D_j}{S_j}$ is twice the periodogram normalized to the best fit.
$T_R$ measures the maximum deviation between the observation and the model and it is used to find the confidence line of any given percentage, $\mathrm{conf}^{k\%}$. This line is a threshold for discerning PSD peaks due to real oscillations on the time series over noise-generated peaks.
In this work we use \(k\%=95\%\).

To determine $\mathrm{conf}^{k\%}$, we begin generating synthetic data based on the posterior distribution of the parameters. This involves creating a large number of parameter vectors \(\theta^{\text{synth}} = \{\theta_1, \dots, \theta_M\}^{\text{synth}}\). 
By using these vectors in our model, we obtain a set of synthetic best fits \(S_j(\theta^{\text{synth}})\), which are expected to follow an exponential distribution relative to the ``real'' PSD.
To simulate synthetic periodogram data, we then multiply each element of these synthetic best-fit vectors by a random draw from a \(\chi^2_2\) distribution, reflecting the expected statistical behavior of the periodogram under the null hypothesis.
The resulting synthetic distribution is then
\begin{equation}
    D_j^{synth} = S_j(\theta^{synth}) X_j / 2 \, .
\end{equation}
Once all the synthetic periodograms are computed, we obtain $R_j^{synth} = 2 D_j^{synth} / S_j(\theta_{mode})$.

For each frequency, we check the value of $R_j^{synth}$ so the p-value $p_B^{k\%}(R_j^{synth})$ is less than the $1 - k\%/100$ confidence value we want to obtain. For example, if we want the 95\% confidence the $p$-values should be less than 0.05. 

Finally the $\mathrm{conf}^{k\%}$ can be obtained from:
\begin{equation}
    \text{conf}_j^{k\%} = R_j^{synth}[p_B < 1 - k\%/100] \cdot S_j(\theta_{mode}) / 2 \, .
\end{equation}
The confidence line follows the same functional form as in Eq.\eqref{eq:backgroundnoise}, but it is not parallel to the best-fit line and has different parameter values. 

We use the Python library $PyMC$ \citep{Abril-Pla_PyMC_a_modern_2023} to implement the MCMC algorithm. 
We used this method to study already reported oscillations in \citet{luna_gong_2018}, and the results obtained were in agreement with the previously reported. The results were stored and used for training the deep learning model presented in the next section. 
The main problem with the Bayesian technique lies in the computing time required. The MCMC and confidence line computations take a minimum of 10 seconds for a single pixel.
To perform this operation on the entire solar disk using GONG \ha images, processing one day of observation would take approximately 230 days on a single CPU core. Even with parallelization techniques, this approach is still too slow. To accelerate this computation, we use deep-learning techniques. An important insight is that the \(k\%\) confidence line can be modeled, as the best-fit line, using Eq.~\eqref{eq:backgroundnoise}. Therefore, this work aims to develop a model capable of obtaining the two sets of parameters that would otherwise result from the MCMC calculation.


\section{Convolutional Neural Network}

\begin{figure*}[!ht]
    \includegraphics[width=0.98\textwidth]{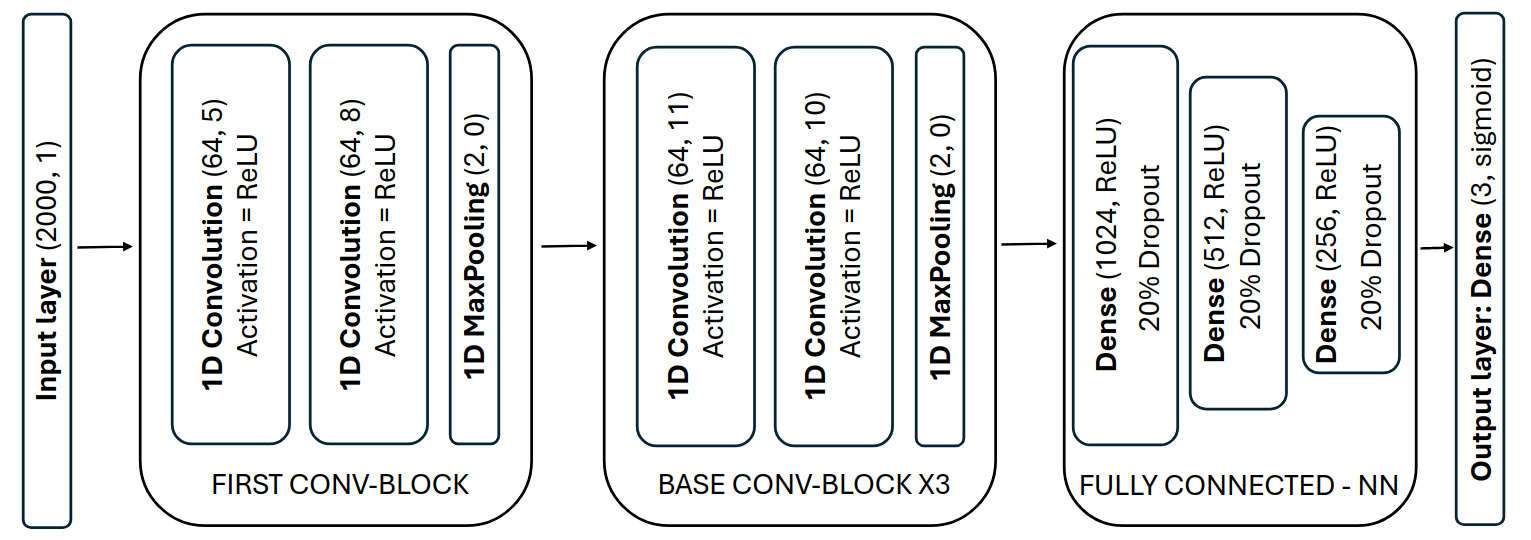}
    \caption{Architecture for the CNN model used in this work. First the input layers that expect the periodogram, the data is then processed by one convolutional block (ConvBlock) with kernel sizes of 5 and 8, proceeded by three identical convolutional blocks with kernel sizes of 11 and 10, each ConvBlock ends with a MaxPooling layer to reduce the dimensionality of our data. Each convolutional layer has the ReLU activation function. The data is then flattened and sent to a small fully-connected neural network with dropouts, which finally predicts the three parameters we are interested in normalized between 0 and 1.}
    \label{fig:architecture}
\end{figure*}

\begin{figure*}[!ht]
    \centering
    \includegraphics[width=\linewidth]{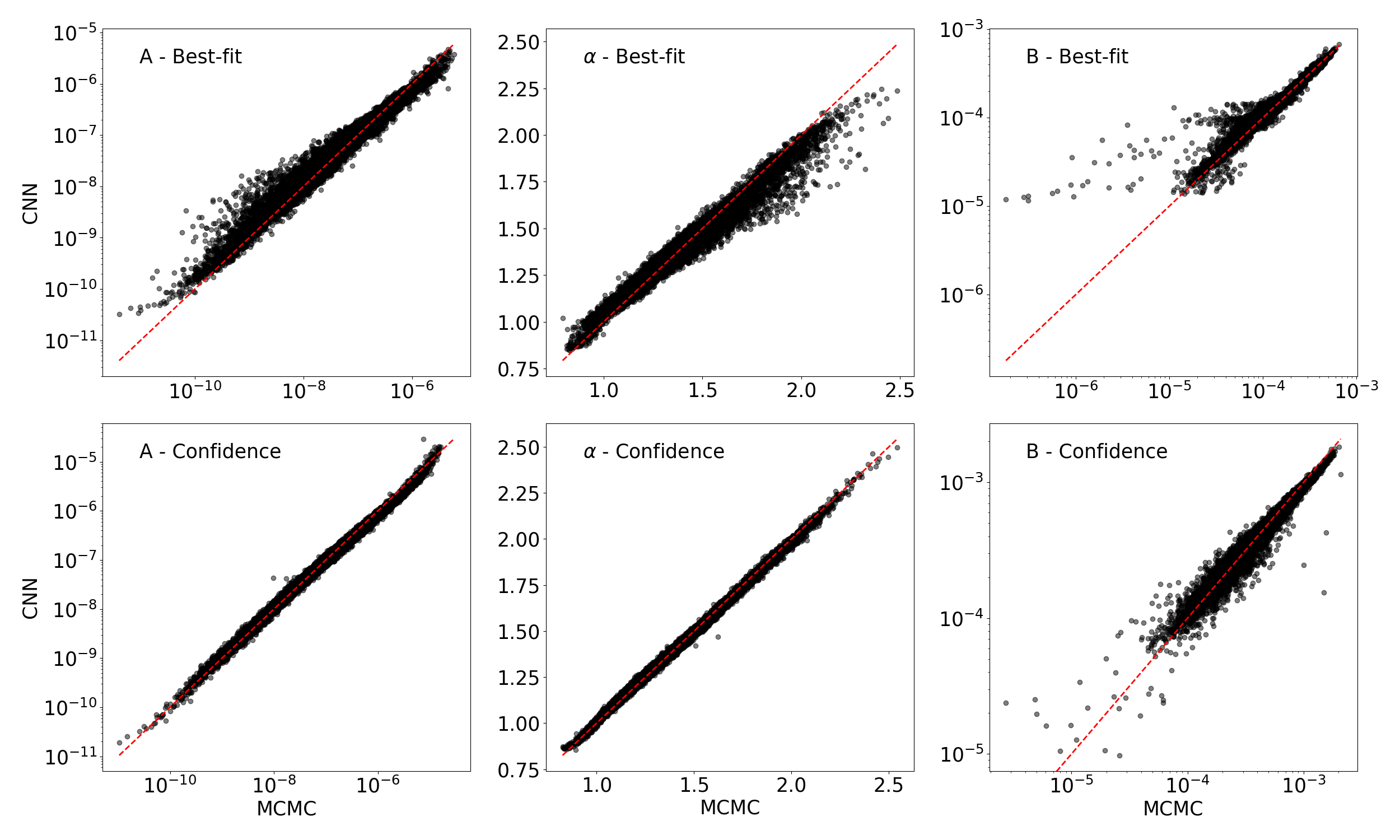}
    \caption{Scatter plots of the expected parameters (MCMC) in the x-axis and the predicted parameters (CNN) in the y-axis for all the periodograms from the testing dataset. The top row is for the best-fit predictions, the bottom row is for the 95\% confidence prediction.}
    \label{fig:params_distributions}
\end{figure*}

CNNs have demonstrated their effectiveness across numerous scientific fields, including solar physics \citep{asensio_ramos_machine_2023}.
In this work, we present a simple CNN model capable of predicting the set of parameters for the best fit and a $k\%$ confidence line of an MCMC algorithm for a given input PSD.
The tools developed and presented in this work are a proof of concept of the capabilities of CNNs for the analysis of solar data, as well as a demonstration of how powerful these tools proved to be for our research purposes.
For this work, we used a server equipped with an NVIDIA RTX A5000 GPU to train our deep learning model, which is capable of estimating the best-fit and confidence line parameters for each pixel in the whole solar disk in approximately 5 minutes once trained.

\subsection{Architecture}
The architecture employed in this study is illustrated in \Cref{fig:architecture}. The model expects as input a periodogram and outputs vector with the values of A, $\alpha$ and B, for either the best-fit or the confidence line. 

The network begins with an input layer that takes the normalized LS-periodogram, followed by four convolutional blocks (ConvBlock). Each ConvBlock consists of two 1D convolutional layers with 64 kernels and ReLU activation, followed by a max-pooling layer with a filter size of 2. The first ConvBlock uses kernel sizes of 5 and 8, while the subsequent blocks utilize kernel sizes of 11 and 10.
Following the convolutional blocks, a small fully connected network with dense layers containing 1024, 512, and 256 units, respectively, are employed, each using ReLU activation. To reduce overfitting, dropout layers with a 20\% drop probability are included after each dense layer.
Finally, the output consists of a dense layer with three units, each corresponding to one of the parameters to be predicted (A, $\alpha$ and B). This layer uses sigmoid activation, as the output data is normalized to values between 0 and 1.
%
%

We used the Adam optimizer \citep{kingma2014adam} and a Mean-Squared Error for training the model. To improve performance a learning rate decay schedule was applied, where for the first ten epochs the learning rate remains constant and after every fifth epoch the learning rate decreased by a factor of 0.7, to prevent over-fitting with this schedule we also incorporated an early stop where we monitor the validation loss.
The training was performed for a maximum of 100 epochs, and we found the most suitable batch size to be 32.

\subsection{Training/Validation data}
This work aims to model the PSD best fit and its $k\%$ confidence line for a given input periodogram.
%
In this study, two datasets has been employed for training the neural network. The first dataset consists of real data analysis using MCMC (Sect. \ref{sec:spectral-analysis}), although its size is relatively small due to the slow nature of the calculations involved, rendering it insufficient for effective training. To overcome this limitation, a second dataset was generated synthetically, which allowed for a significant increase in the amount of available information for training purposes.
The synthetic data was constructed using real data as a reference. Specifically, we utilized the posterior distributions derived from the MCMC method and parameterized them. Random noise was then introduced to these parameters to introduce slight variations.
%
From these modified posterior distributions, we derived the best-fit parameters and confidence intervals using the methods described in Sect. \ref{sec:spectral-analysis}.
The corresponding periodograms were constructed from the best-fit parameters and the null hypothesis based on the Bayesian approach (see also \citet{vaughan_bayesian_2010} for details in the appendices).
The best-fit line is multiplied by random values sampled from an exponential distribution to mimic the same statistical behavior as the real data. 
We generate a total of $2 \times 10^6$ synthetic periodograms, each associated with a best-fit and confidence interval parameter set. The data was divided using a standard 80-20 split for training and validation. Input features were scaled to a range of 0 to 1 by default, and outputs were normalized using a logarithmic transformation followed by Min-Max scaling to further constrain values between 0 and 1.

\subsection{Model testing}
To assess the reliability and performance of the CNN model, we compare its results with those obtained from observational data.
From GONG data, we have created a testing dataset of $10^5$ periodograms selected randomly from different observation days and the  best-fit and confidence curves are computed using the MCMC techniques.

Figure \ref{fig:params_distributions} shows a comparison of the predicted distributions of expected parameters (MCMC) in the $x$-axis and predicted parameters (CNN) in the $y$-axis. The top row is for the Best-Fit model and the bottom row is for the 95\% confidence model. In an ideal case, all predictions should lie on top of the red-dashed diagonal line, so for any given input both methods (MCMC and CNN) should predict exactly the same values. The narrower the distribution is around the red-dashed line the better the model performs.
The distributions for the Best-Fit model are wider than the 95\% confidence model, indicating poorer performance. Both models struggle with the predictions of the B parameter, showing the most outliers and wide distributions.
Using the predicted parameters we can compute the lines given by Eq. \eqref{eq:backgroundnoise}.

\begin{figure}[!h]
        \includegraphics[width=0.5\textwidth]{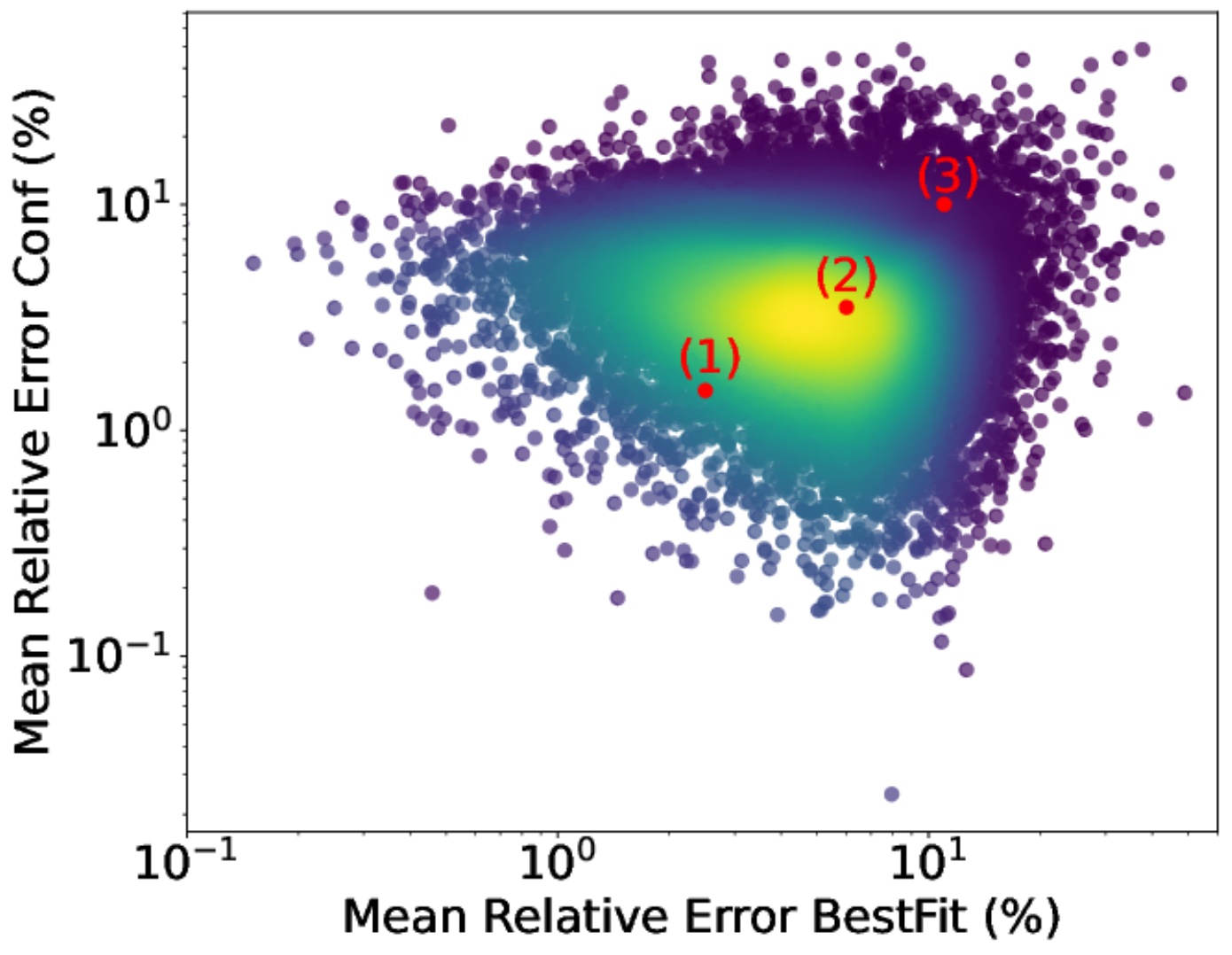}
    \caption{2D histogram representing the percentual relative errors found in the testing dataset, the horizontal axis is for the best-fit errors while the vertical axis is for the confidence line error. Colors represent the density of points, with yellow indicating the highest density zones and purple the smaller ones. The mean values of the distributions are
    $\mu_{confidence} = 4\%$ and $\mu_{best-fit} = 8\%$. The red dots indicate samples from the different regions of the distribution that will be visualized in the next figure.}
    \label{fig:test_error}
\end{figure}
Figure \ref{fig:test_error} shows the relative errors between both the CNN and MCMC model predictions for the best-fit (horizontal axis) and confidence lines (vertical axis). 
%
The mean relative error for the best-fit line is approximately 8\%, while for the confidence interval, it is 4\%. These values indicate that, on average, errors for the best-fit line are higher than those for the 95\% confidence line errors, yet remain within acceptable limits.

\begin{figure}[!htpb]
    \centering
    \includegraphics[width=\linewidth]{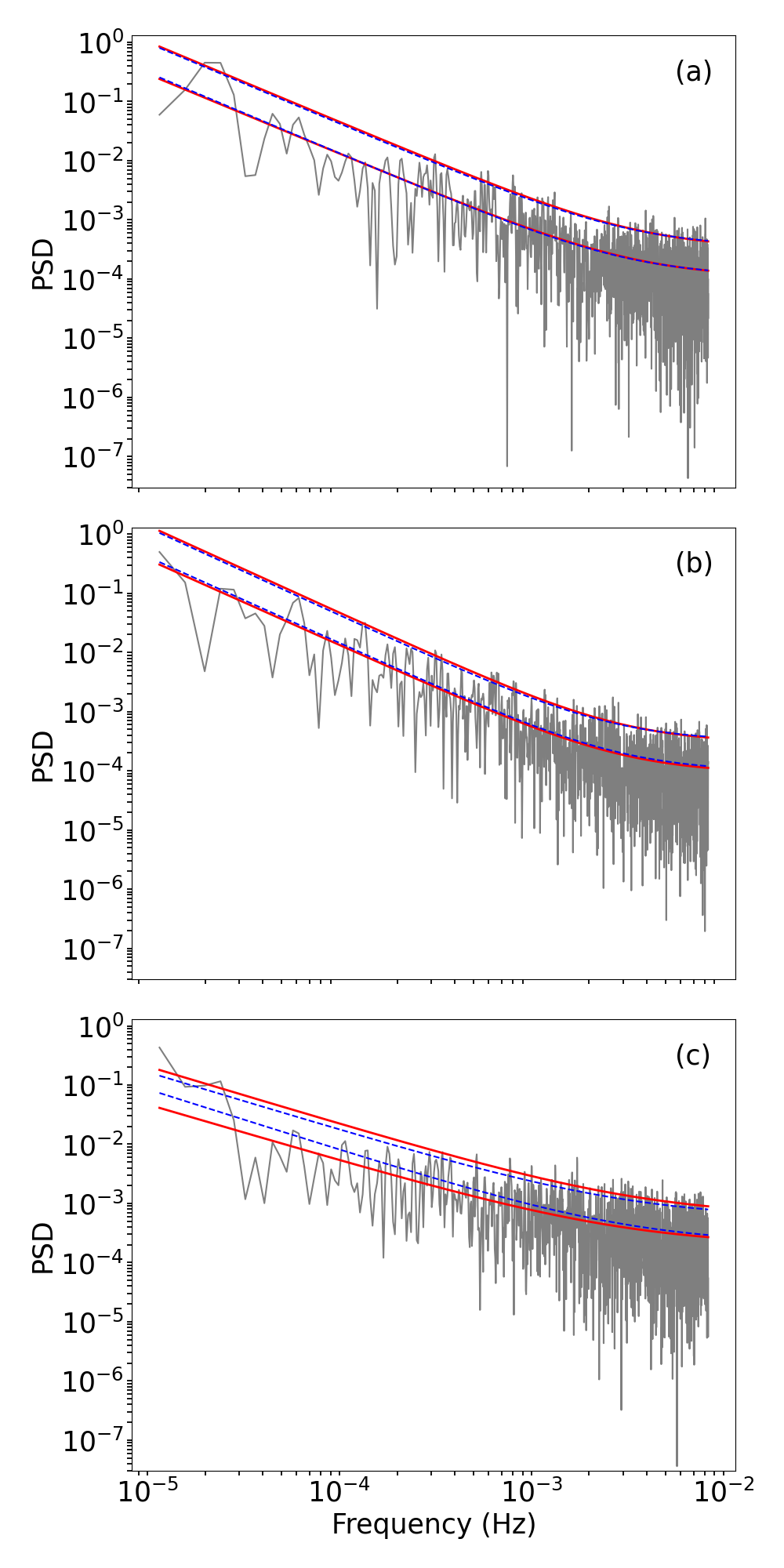}
    \caption{Periodograms with the predicted best-fit and 95\% confidence curves from the test dataset, highlighted in red dots (1), (2), (3) in \cref{fig:test_error}. (a) refers to region (1) with the smaller errors, (b) to region (2) with average errors, and (c) to region (3) with the highest errors. }
    \label{fig:specific_cases}
\end{figure}

Figure \ref{fig:specific_cases} shows three examples of periodograms, together with their corresponding best-fit and confidence curves. These examples are selected from the test dataset and are marked with red dots on \cref{fig:test_error}.
%
In \cref{fig:specific_cases}a we plot case (1) with a relative error of 2\% in both the best fit and confidence line.
%
%
Both sets of curves almost overlap in all the frequency domain considered. The second example in \cref{fig:specific_cases}b displays a greater discrepancy between the CNN and MCMC results. For the best-fit and confidence curves, the relative errors are 6\% and 5\%, respectively.
Despite a slight separation between the curves, they are nearly identical across most of the frequency range.
Finally, in \cref{fig:specific_cases}c, we show a case (3), where the errors exceed 10\% on average. 
This is the worst case shown in \cref{fig:specific_cases} with 11\% error on the best-fit and 10\% of the confidence line.
Both sets of curves exhibit the largest discrepancies, with a greater deviation observed in the best-fit case. A detailed inspection reveals that it is the MCMC curve that deviates, while the model result performs better. In contrast, the confidence line shows more consistent results across both methods.
As seen in \cref{fig:test_error}, the number of cases like (3), where the errors are high, are not frequent. Specifically, for our testing dataset, we found that only 2\% of the periodograms have on average more than 10\% error.

\section{Results: Study of different events}\label{sec:results}
In this section, we apply the neural network model to three different days of GONG data. The first two days include events already reported in the \citet{luna_gong_2018} catalog (hereafter referred to as ``the catalog''). In this way, we evaluate the validity of the model. The third case shows new oscillations where we prove the ability of the model to detect new events. 
%
\subsection{1 January 2014}
\begin{figure*}[!htbp]
    \centering
    \includegraphics[height=0.9\textheight]{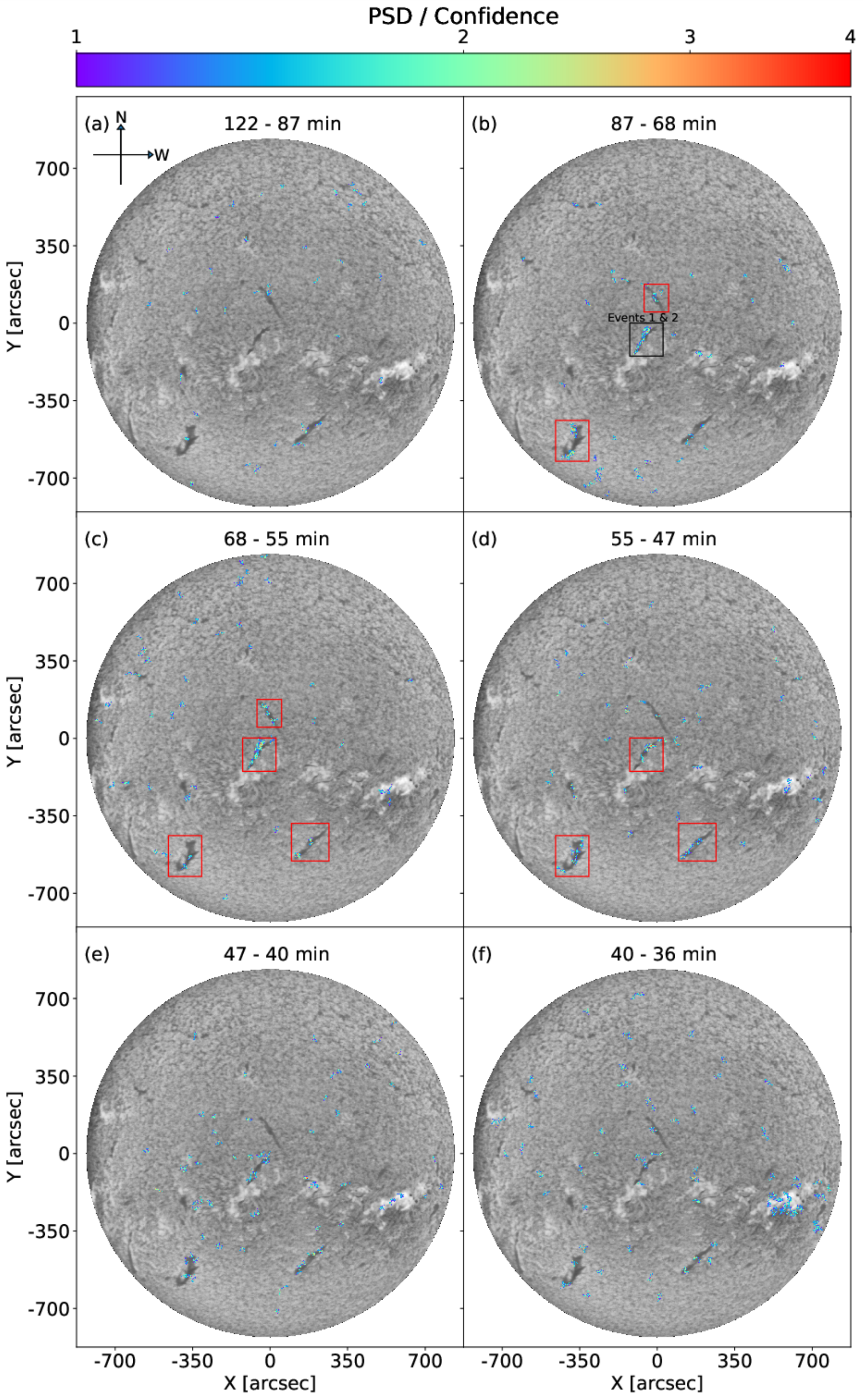}
    \caption{Ratio of PSD over the predicted confidence line from the CNN model for the 1 January of 2014 over the background \ha intensity map (the \ha maps have artificially increased contrast for clearer visualization of the dark structures). From (a) to (f) the predictions have been binned in frequency with an increment of 0.055 mHz. We show the approximate period ranges associated with each frequency bin. The black boxes represent events reported in \citet{luna_gong_2018}, within the respective period range. The red boxes represent newly detected events we believe are related to filament oscillations that will be further discussed in the text.}
    \label{fig:20140101_collage_low_freq}
\end{figure*}

Figure \ref{fig:20140101_collage_low_freq} shows the oscillation analysis of the full disc of 1 January 2014. 
To construct these figures, the ratio of the PSD to the 95\% confidence curve is calculated, considering only the cases of positive detection, i.e., where $\mathrm{PSD}/\mathrm{conf}^{95\%} \ge 1$.
Additionally, the frequency range is divided into bins of 0.055 mHz, spanning from 0.136 to 0.461 mHz. The panels display the maximum ratio between the PSD and the confidence level within each frequency bin. 
%
%
This ratio is overplotted on the \ha images (which have increased contrast for better visualization of the filaments). This allows us to determine whether the oscillations are associated with the filaments.
%
%
In the catalog, two different events have been reported (event numbers 1 and 2) both in the same filament placed almost at the center of the disk $(x, y) \approx (-18, -76)$ arcsecs shown in \cref{fig:20140101_collage_low_freq}b with a black box. The first event was reported to have a period of 75.2 $\pm$ 1.0 minutes, and the second event with a period of 71.2 $\pm$ 2.8 minutes.

Figure \ref{fig:20140101_collage_low_freq}a shows the ratio in the first bin, between 87-122 minutes. 
Numerous smaller, dispersed regions are present, many of which are not on filaments. In panel \ref{fig:20140101_collage_low_freq}b, between the periods 68 to 87 minutes there is considerable power over the three filaments marked with boxes.
The filament enclosed with a black box corresponds to events 1 and 2 with the period in this frequency range.
This shows that the findings of the CNN model agrees with the results of the catalog. 
Additionally, there is also power around two other filaments marked with red boxes. These oscillations were not reported in the catalog.
Similarly to panel \ref{fig:20140101_collage_low_freq}a we see small areas detected scattered around the disk. 
Figure \ref{fig:20140101_collage_low_freq}c shows the power distribution between 55 and 68 minutes. 
In this panel, we see that the central filament also has power in this period range. This indicates that the oscillations associated with events 1 and 2 of the catalog are distributed in two frequency bins. As we see below, this is because the PSD has a Gaussian distribution containing at least these two bins.
%
%
%
Additionally, some power is visible in the two other filaments shown in \ref{fig:20140101_collage_low_freq}b, along with a fourth filament enclosed in a red box in the south-central region of the disc centered at approximately $(x, y) = (400, -400)$ arcsecs.
For the northernmost highlighted filament, the PSD is distributed around the filament structure and displays multiple frequencies exceeding the confidence threshold. The most significant contributions come from the two frequency bins corresponding to period ranges of 68–87 minutes and 55–68 minutes. 
%
%
In the following panels, \cref{fig:20140101_collage_low_freq}d to \cref{fig:20140101_collage_low_freq}e, some power also appears over the filaments. However, across the disc, it is seen in the form of small regions. Similarly, in \cref{fig:20140101_collage_low_freq}f, numerous small regions are scattered across the disc. Notably, there is significant power in the 
western solar plage, which is attributed to the activity in this active region (AR).
Small filaments also exhibit PSD values above the confidence threshold, marking the detection of additional oscillations in various frequency bins. While a detailed analysis of these smaller filaments is beyond the scope of this paper, they represent an important dataset for future statistical research.
To reduce the number of figures, we omit the panels similar to \cref{fig:20140101_collage_low_freq} corresponding for periods shorter than 36 minutes. In those panels, numerous small regions of relatively low power appear scattered across the disc.
%
\begin{figure}[!htbp]
    \centering
    \includegraphics[width=0.45\textwidth]{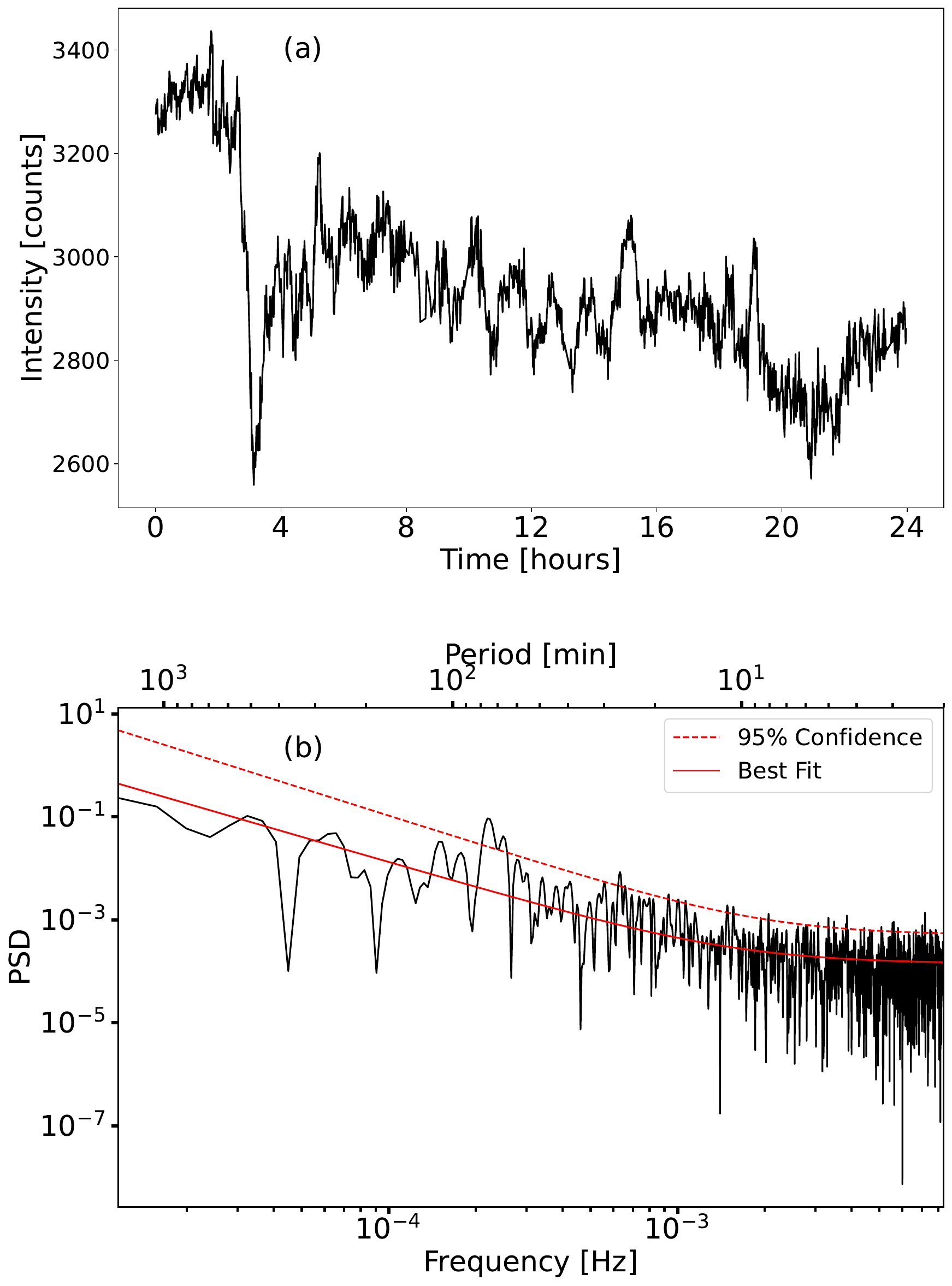}
    \caption{In (a), a plot showing the \ha intensity as a function of time during the 24 hours of 1st January 2024. The intensity is from a pixel inside the black box from \cref{fig:20140101_collage_low_freq}b. In (b) the corresponding PSD and the predicted best fit (solid red line) and 95\% confidence line (dashed red line) are shown.}
    \label{fig:20140101_psd_center_filament}
\end{figure}

In \cref{fig:20140101_psd_center_filament}a we show the \ha intensity in a pixel within the region marked with a black box in \cref{fig:20140101_collage_low_freq}b. 
The best fit and confidence lines constructed with our neural network model are also plotted in \cref{fig:20140101_psd_center_filament}b.
%
This figure is similar to the results presented in Figure 5 by \citet{luna_automatic_2022}. It is important to note that in their figure, both lines -the confidence line and the fit- are parallel. In contrast, our case clearly shows non-parallel lines due to the application of a more advanced technique (Sect. \ref{sec:spectral-analysis}).
%

%
In \cref{fig:20140101_psd_center_filament}a periodic fluctuations are clearly present from 5:00 to 16:00 UT approximately. These periodic fluctuations are associated with the passage of the filament over the pixel that produces periodic occultations of the chromosphere.
In the PSD  shown in \cref{fig:20140101_psd_center_filament}b, the most significant peak is centered at 73 minutes with a value of approximately 3 times the confidence level, well above the threshold. This detection is in agreement with the results found in the catalog. This peak falls into two different frequency bins showing power in both panels \cref{fig:20140101_collage_low_freq}b and \cref{fig:20140101_collage_low_freq}c. There is also a secondary peak centered at 64 minutes, with values almost 2 times bigger than the confidence threshold.
In addition, other peaks in different frequencies are also over the confidence threshold as in a peak around 30 minutes, and multiple presented below 10 minutes.
However, these frequencies are more localized in space as seen in Figs. \ref{fig:20140101_collage_low_freq}d to \ref{fig:20140101_collage_low_freq}f where small areas appear over the filament.
%
This indicates that this type of filament oscillation is multi-periodic.
This could be associated either with the existence of overtones and/or with different parts of the filament oscillating with different periodicities along the line of sight.

\subsection{13 February 2014}
Figure \ref{fig:20140213_collage_low_freq} shows the oscillatory analyses performed for the 13th of February of 2014. In the catalog, there are two events reported on this day, events 62 and 63. Event 62 was reported for a filament centered around $(x, y) = (600, 239)$ arcsecs with a period of oscillation of $ 47.6 \pm 0.6$ minutes. Event 63 was reported in a big filament in the 
southwest of the disk at approximately $ (x, y) = (708, -382)$ arcsecs, with a period of $103.3 \pm 0.4$ minutes. 
Figure \ref{fig:20140213_collage_low_freq} panels represent the same frequency bins that we considered previously. Similarly to the previous case, many detection zones appear on the different bins scattered around the disk. In \cref{fig:20140213_collage_low_freq}a, we see considerable PSD around the filament in the 
southwest of the disk, highlighted with a black box. This corresponds to event 63 from the catalog.
Similar results can be seen in \cref{fig:20140213_collage_low_freq}b.
Panel \ref{fig:20140213_collage_low_freq}c, like in all the other considered bins has scattered PSD through the disk, but a remarkable detection can be seen in the active region placed at $(x, y) = (300, -100)$ arcsecs, highlighted with a red box.
In \cref{fig:20140213_collage_low_freq}d, regions in the filament where event 63 was reported are detected. Additionally, another filament north of this region shows a strong detection. Both filaments are highlighted with a black box. This detection, absent in the previous bins, corresponds to event 62 from the catalog.
Like in the previous case studied, the detection peak of this event falls between 2 of the bins used, so there are detection zones in panel (e) in similar zones to panel (d) for both events reported in the catalog.
Finally in \cref{fig:20140213_collage_low_freq}e and \cref{fig:20140213_collage_low_freq}f we can see again the concentration of PSD in the active region.
%

\begin{figure}[!htbp]
    \centering
    \includegraphics[width=0.45\textwidth]{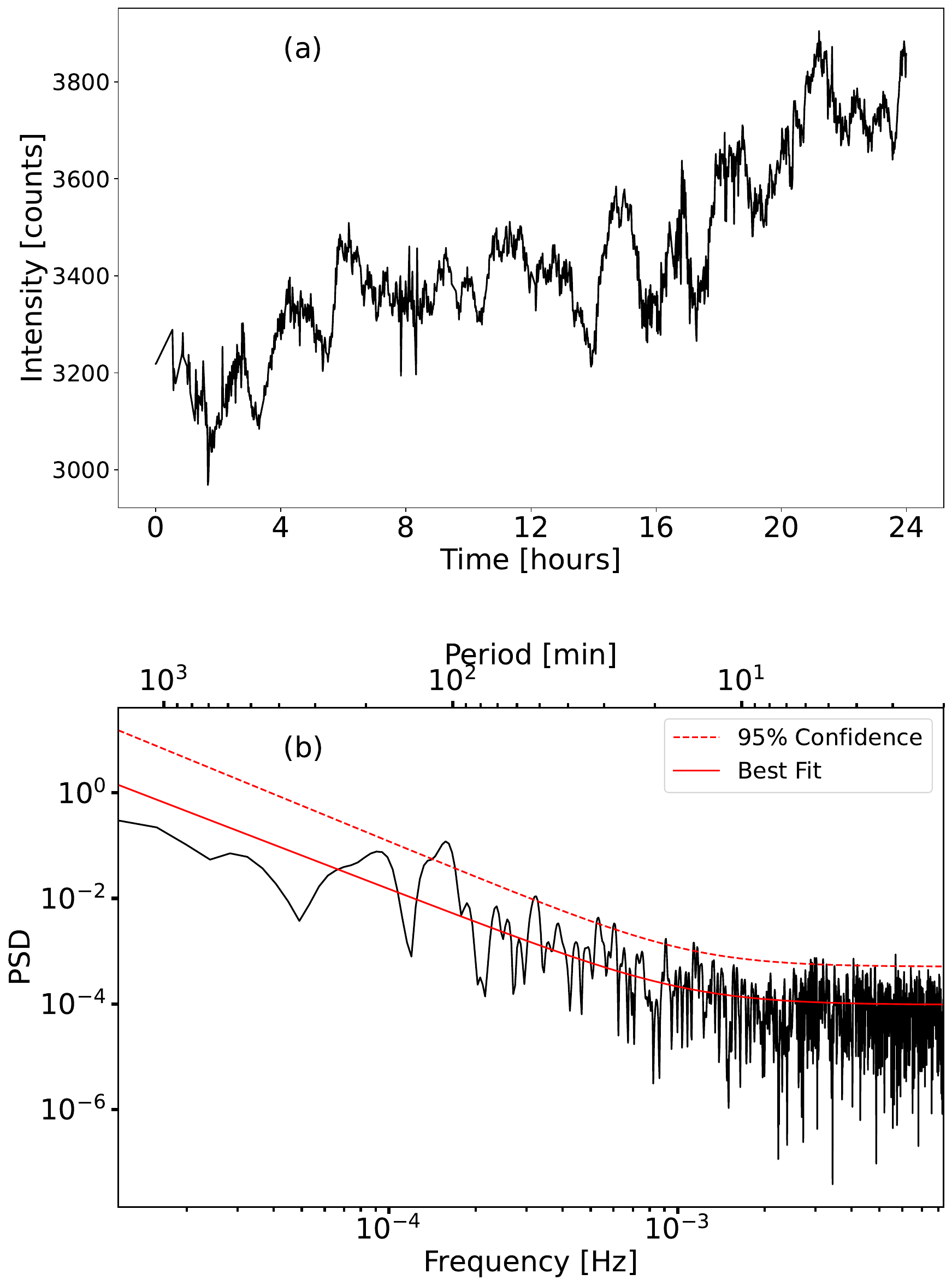}
    \caption{Same as \cref{fig:20140101_psd_center_filament} but for a pixel from the black box highlighting the event 63 from the catalog.}
    \label{fig:20140213_time_psd}
\end{figure}

Figure \ref{fig:20140213_time_psd}a shows the temporal evolution of a pixel where the ratio $\mathrm{PSD}/\mathrm{conf}^{95\%}$ was the highest for the event 63 and in \cref{fig:20140213_time_psd}b the corresponding periodogram with CNN predicted best-fit and 95\% confidence curves. 
The main peak is centered around 100 minutes, but there is a secondary peak at $\sim$ 50 minutes and more between 40 to 30 minutes. The position of the main peak is in agreement with \citet{luna_automatic_2022} Figure 6(c), but in their figure, those secondary peaks are not visible. 
The presence of these peaks needs to be treated with caution. \citet{luna_automatic_2022} showed the existence of a secondary period associated with the filament crossing its equilibrium position, with a period that is half of the primary oscillation period. The PSD at this period is concentrated over the filament. 
This does not imply that a zone detected at half the period of the stronger signal will always be an artifact of this type; however, some of these detections should be examined with caution.
In future statistical studies, these cases will be addressed using alternative techniques to avoid reporting events that result from this method limitation.

\begin{figure}[!htbp]
    \centering
    \includegraphics[width=0.45\textwidth]{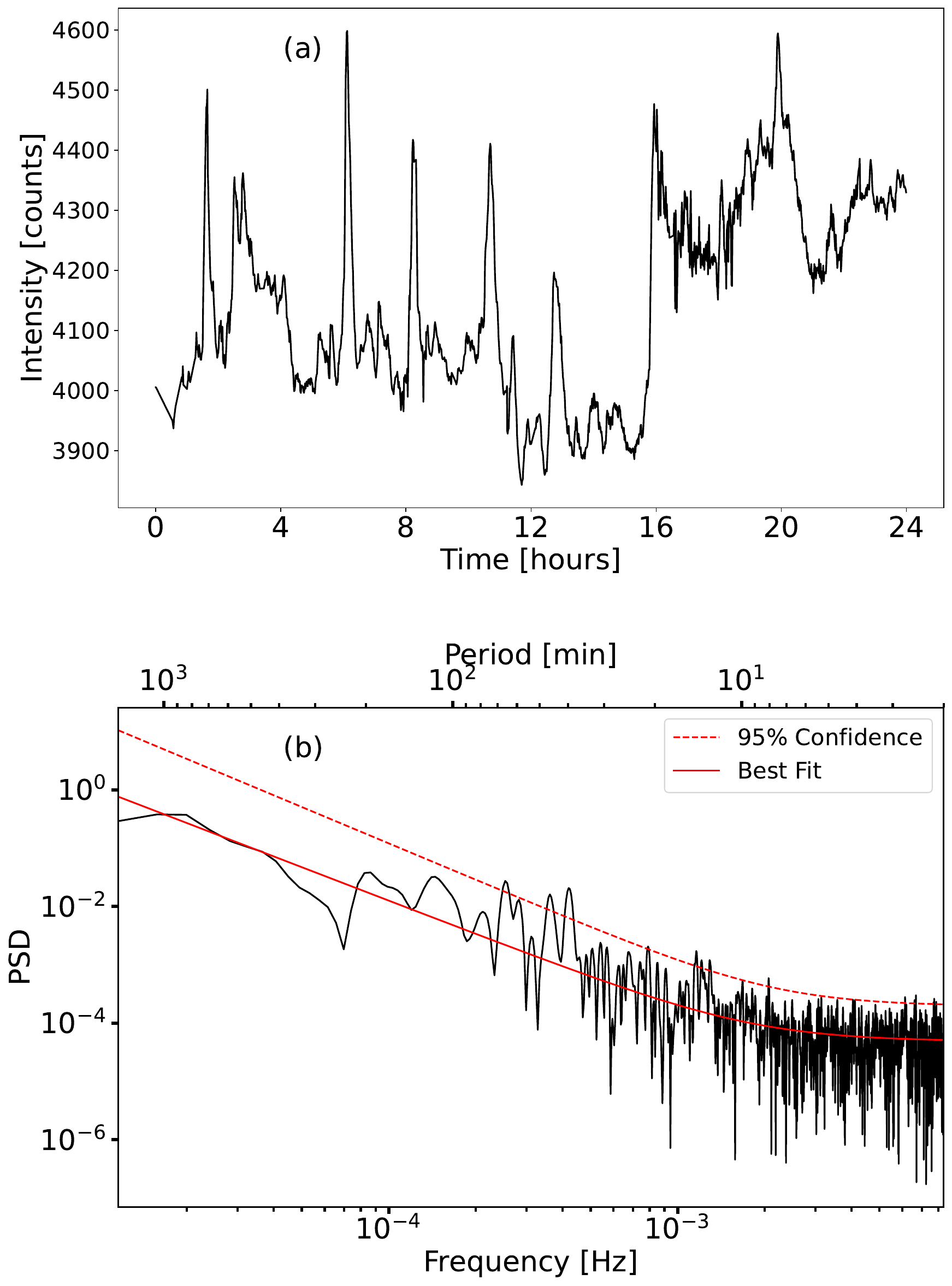}
    \caption{Same information as in \cref{fig:20140101_psd_center_filament} but for the average of the active region seen at  $(x, y) = (300, -100)$ from the 13th of February 2014.}
    \label{fig:20140213_active_region}
\end{figure}

Finally, \cref{fig:20140213_active_region} shows the average temporal evolution of the active region with the associated periodogram.
As shown in panel \ref{fig:20140213_active_region}a, the active region exhibits a repetitive sequence of 
intensity peaks associated with C and M-class flares 
according to Heliophysics Events Knowledgebase \citep[HEK;][]{Hurlburt2012} database. In panel \ref{fig:20140213_active_region}b, we observe that the positions of the three main peaks align with the visual results from \cref{fig:20140213_collage_low_freq}c, \ref{fig:20140213_collage_low_freq}e, and \ref{fig:20140213_collage_low_freq}f. The primary peak is centered at 39 minutes, followed by another at 47 minutes, and the final one at 66 minutes.
These results demonstrate that our techniques can detect other periodic events within solar chromosphere observations and are not limited to filament oscillations alone, revealing the effectiveness of this machine-learning approach.

\subsection{22 October 2024}
Finally, we analyze the data for 22 October 2024. This day has not been previously studied and has been selected because a large number of filaments appear. As in the previous cases, \cref{fig:20241022_collage_low_freq} shows the PSD in the different bins obtained from the CNN predictions.
%

In Figs. \ref{fig:20241022_collage_low_freq}a to \ref{fig:20241022_collage_low_freq}f, the PSD is predominantly scattered across the solar disk. Notably, there are specific regions where the PSD exhibits concentrations near filaments; these areas have been highlighted with black boxes in the corresponding frequency bins. These concentrations suggest localized oscillatory phenomena associated with filament structures.
Specifically, in Figs. \ref{fig:20241022_collage_low_freq}a and \ref{fig:20241022_collage_low_freq}b, we detect an event at the small filament located at approximately \((x, y) \approx (-320, 80)\) arcsec with periods exceeding 70 minutes. 
%
The subsequent panels highlight three additional filaments that demonstrate PSD concentrations in various frequency bins, indicating the presence of multiple periodicities. 
The easternmost filament, situated at \((x, y) \approx (-400, -200)\) arcsec, shows an event triggered by a 
C-class flare 
(according HEK database) at 
around 12:10:00 UT emanating from a nearby active region. This flare induces movement in the eastern end of the filament, resulting in oscillations of the entire structure. The complexity of these oscillations is evident, with multiple periodicities detected across different regions of the filament, as depicted in Figs. \ref{fig:20241022_collage_low_freq}c to \ref{fig:20241022_collage_low_freq}f. These observations highlight the intricate coupling between flares and filament dynamics.
Another significant filament oscillation is observed at coordinates \((x, y) \approx (100, -400)\) arcsec, as shown in Figs. \ref{fig:20241022_collage_low_freq}d and \ref{fig:20241022_collage_low_freq}e. This event corresponds to a periodic movement of the filament without an apparent external trigger, as no immediate cause is evident from the visual inspection of its temporal evolution. A dedicated study would be necessary to understand the trigger of this oscillation.
Finally, the westernmost event involves the filament located at \((x, y) \approx (680, 10)\) arcsec. The periodicities associated with this filament are displayed in \cref{fig:20241022_collage_low_freq} panels (c), (d), and (f). This filament's oscillation is initiated by a 
C-class flare event occurring around 15:20:00 UT from a small active region situated to the south. The flare induces motion in the filament, causing it to oscillate with different periods. Shortly after the initial flare, a subsequent 
C-class flare occurs leading to its eruption and disappearance from the \ha data. 
%
%
Our CNN model effectively detects and analyzes solar filament oscillations, as demonstrated by the successful identification of oscillatory events on 22 October 2024. By automating the detection and spectral analysis processes, the model enhances the efficiency and comprehensiveness of studying filament dynamics, contributing to a deeper understanding of the physical mechanisms involved.

\section{Summary and conclusions}\label{sec:discussion}

In this study, we have investigated the detection of oscillations in images using machine learning techniques. Following the approach of \citet{luna_automatic_2022}, we applied a spectral technique that analyzes periodic intensity fluctuations in each pixel of GONG \ha data.
%
%
Firstly, data cubes are generated by stacking the images from a day collected by the various GONG telescopes. The images to be used are automatically selected from all the telescopes in the network, with low-quality images discarded, the intensity of the images from different telescopes regularized, and differential rotation compensated. This results in a data cube that covers an entire day. The temporal cadence is one minute, so ideally, there would be 1440 images per day. However, there are not always images available, or they may have been eliminated during our selection process. This can result in gaps in the time sequence.
%
%
For each of the pixels in the data cube, the PSD is calculated using the Lomb-Scargle periodogram.
As usual in these spectral techniques, we consider detection to occur when a PSD peak exceeds a certain threshold above the background noise.
The background noise is well-modelled by a combination of red and white noise.
%
%

The background noise signal is determined by fitting the PSD to a model, which in our case is a combination of red and white noise. Additionally, several techniques are available for establishing the confidence line. 
%
%
%
However, it seems that the most appropriate in the presence of red noise in addition to white noise relies on Bayesian statistics and MCMC techniques \citep{vaughan_bayesian_2010}, as recommended by \citet[][see Section 7.4.2.4]{vanderplas_understanding_2018}. While these methods are powerful for analyzing red-noise periodograms, they proved too slow for our research purposes. To address this, we developed a CNN with training data obtained from the MCMC method. The CNN predicts the same results as the MCMC method but with drastically reduced computing time. This advancement enhances the efficiency of detecting filament oscillations in large datasets, facilitating more comprehensive studies of these solar phenomena.
We performed a thorough error analysis of our deep learning model, obtaining margins of error of 5\% on average on the predicted curves, which are acceptable for our research objectives. The model's capabilities were validated by comparing its results with spectral analyses previously reported in \citet{luna_automatic_2022} and with classical methods from \citet{luna_gong_2018}. The CNN demonstrated high accuracy in predicting oscillation periods, confirming its reliability for this application.
Furthermore, we applied our CNN to a new set of observations from a day not previously studied. Using our model, we identified various types of filament oscillation events, showcasing CNN's ability to detect and classify oscillations in unseen data. This not only highlights the model's generalization capabilities but also contributes new findings to the study of solar filament dynamics.

In conclusion, this study proves the effectiveness of machine learning techniques, through the application of a CNN, in detecting and analyzing oscillations in GONG \ha images. 
This technique yields results comparable to Bayesian methods while significantly reducing computation time by several tens of thousands of times.
This will enable us to analyze vast amounts of information to detect oscillations in the GONG data.


In this paper, we analyse oscillations in the \ha data without distinguishing whether they belong to filaments or not. We have seen that periodic flares can be found and also many regions outside filaments. For a massive study, segmentation techniques will be used to determine the location of oscillations relative to filaments. This will be the subject of a follow-up study.
Although the technique presented in the current work has been implemented in GONG data, it could also be applied to other datasets, such as the EUV images from the SDO. This would enable the detection of oscillations not only in filaments but also in coronal loops and other structures. In future work, we will apply this technique to SDO data.

\section*{Data availability}
The code is publicly available in the following repository:
\url{https://github.com/GuillemCastello/PeriodogramCNN}

\begin{acknowledgements}
This publication is part of the I+D+i project PID2023-147708NB-I00 funded by MICIU/AEI/10.13039/501100011033/  and  by  FEDER, EU.
It has been also part of the R+D+i project PID2020-112791GB-I00, financed by MCIN/AEI/10.13039/501100011033.
M. Luna acknowledges support through the Ramón y Cajal fellowship RYC2018-026129I from the Spanish Ministry of Science and Innovation, the Spanish National Research Agency (Agencia Estatal de Investigación), the European Social Fund through Operational Program FSE 2014 of Employment, Education and Training and the Universitat de les Illes Balears. The authors also thank Prof Ramón Oliver for his helpful comments and suggestions.
\end{acknowledgements}
%
\bibliographystyle{aa.bst} 
\bibliography{bibliography.bib} 
%



 \begin{appendix}
\onecolumn
 \section{Results from 13 February 2014}
\begin{figure}[!htpb]
    \centering
    \includegraphics[height=0.85\textheight]{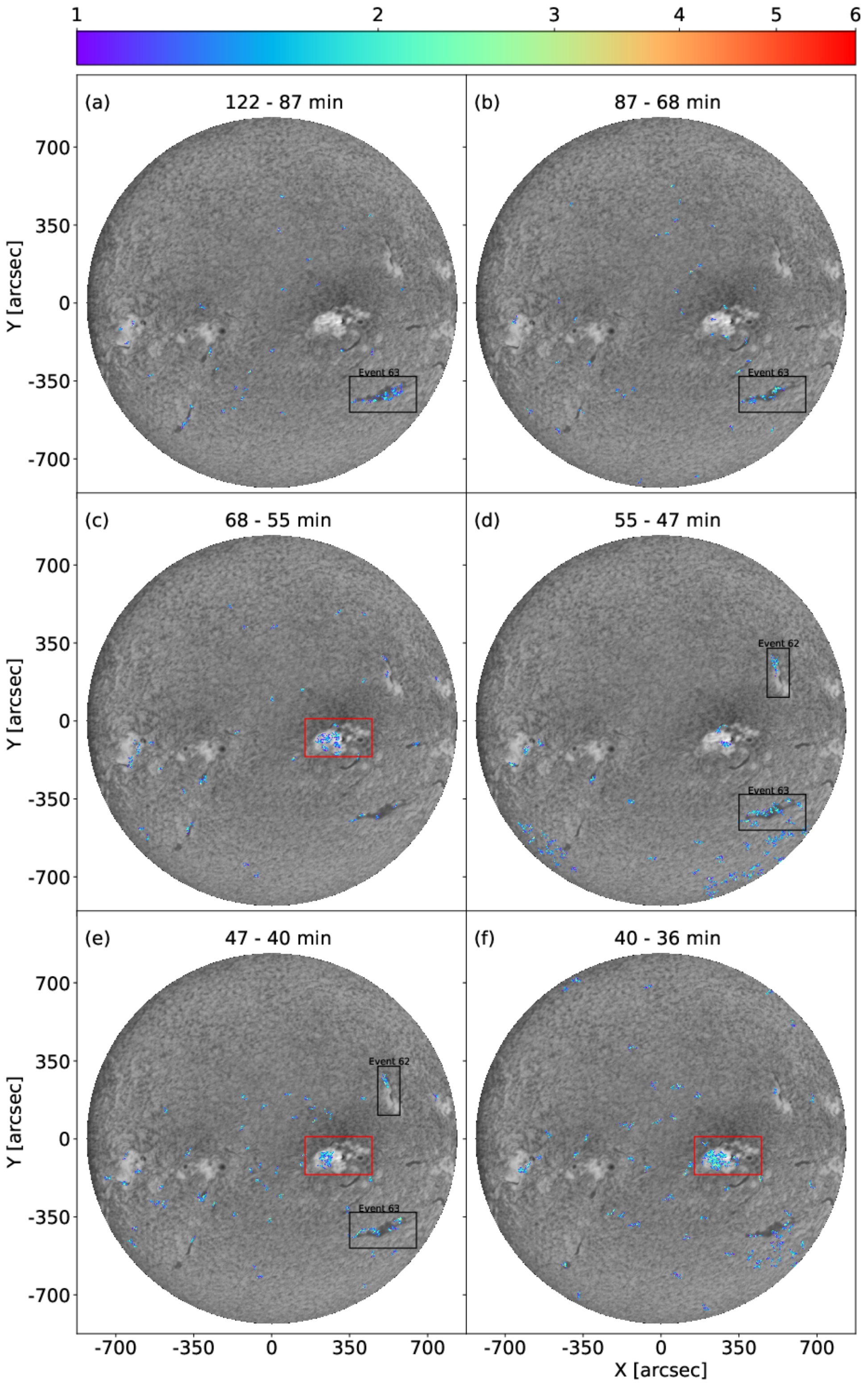}
    \caption{Same as in \cref{fig:20140101_collage_low_freq} for 13 February 2014.}
    \label{fig:20140213_collage_low_freq}
\end{figure}
\FloatBarrier

\newpage

\section{Results from 22 October 2024}
\begin{figure}[!htpb]
    \centering
    \includegraphics[height=0.85\textheight]{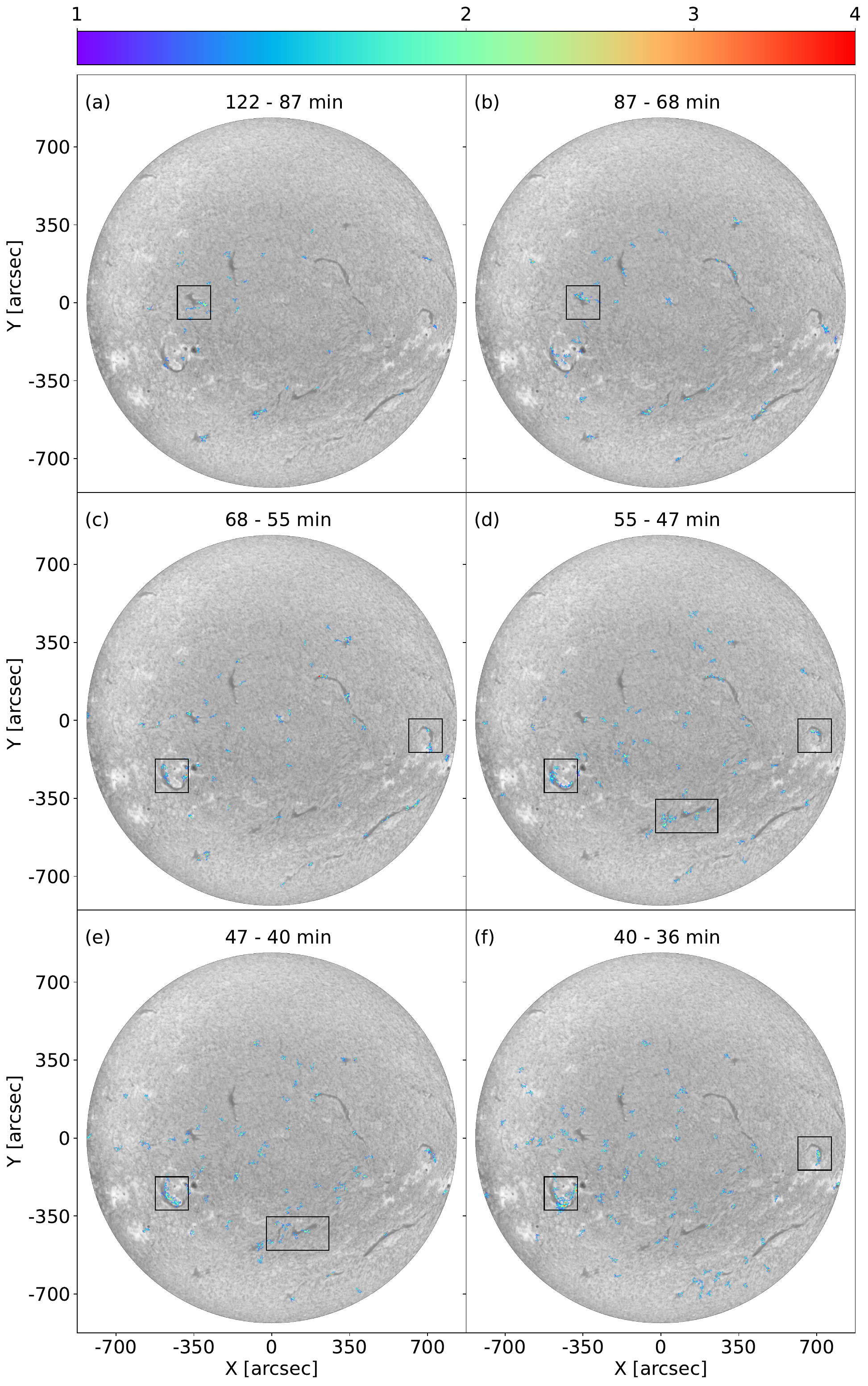}
    \caption{Same as in \cref{fig:20140101_collage_low_freq} for 22 October 2024. 
    Black boxes highlight regions of interest identified in the analysis and discussed in the text.}
    \label{fig:20241022_collage_low_freq}
\end{figure}
\FloatBarrier
\clearpage

\end{appendix}
\end{document}